\documentclass{emulateapj}
\usepackage{epsfig}

\newcommand{\be}{\begin{equation}}
\newcommand{\ee}{\end{equation}}
\newcommand{\bea}{\begin{eqnarray}}
\newcommand{\eea}{\end{eqnarray}}
\newcommand{\MR}{{\it MR}}
\newcommand{\MP}{{\it MP}}

\usepackage{amsmath}
\usepackage{graphicx}

\usepackage[usenames]{color}

\begin{document}

\title{Dynamical models for the Sculptor dwarf spheroidal in a $\Lambda$CDM universe}
\author{Louis E. Strigari}
\affil{Mitchell Institute for Fundamental Physics and Astronomy, 
Dep. of Physics and Astronomy, Texas A \& M Univ., 
College Station, TX 77843, USA}
\author{Carlos S. Frenk}
\affil{Institute for Computational Cosmology, Dep. of Physics,
    Univ. of Durham, South Road, Durham  DH1 3LE, UK} 
\author{Simon D. M. White}
\affil{Max-Planck-Institut f\"{u}r Astrophysik,
Karl-Schwarzschild-Stra\ss{}e 1, 85740 Garching bei M\"{u}nchen,
Germany}

\title{Dynamical models for the Sculptor dwarf spheroidal in a $\Lambda$CDM universe}

\begin{abstract}
  The Sculptor dwarf spheroidal galaxy appears to contain two distinct stellar
  populations of differing metallicity.  Several
  authors have argued that in order for these two populations to
  reside in the same gravitational potential, the dark matter halo
  must have a core similar to that observed in the stellar count
  profile.  This would exclude cuspy Navarro-Frenk-White (NFW)
  density profiles of the kind predicted for halos and subhalos by dark matter
  only simulations of the $\Lambda$CDM 
  cosmological model. We present a new theoretical framework to
  analyze stellar count and velocity observations in a self-consistent
  manner based on separable models, $f(E,J)=g(J)h(E)$, for the
  distribution function of an equilibrium spherical system.  We use
  this machinery to analyze available photometric and kinematic data
  for the two stellar populations in Sculptor. We find, contrary to
  some previous claims, that the data are consistent with populations
  in equilibrium within an NFW dark matter potential with structural
  parameters in the range expected in $\Lambda$CDM; we find no statistical preference for a potential with a core. 
  Our models allow a maximum circular velocity for Sculptor between 20 and 35 km/s. We
  discuss why some previous authors came to a different conclusion.
\end{abstract}

\maketitle

\section{Introduction} 

$\Lambda$CDM has emerged as the standard model of cosmic structure
largely because of its successful predictions for the temperature
anisotropies of the cosmic microwave background radiation, the power
spectrum of the large-scale distribution of galaxies, the structure of the main body of 
galaxy and galaxy cluster halos, as inferred from weak gravitational lensing, and the broad
features of the galaxy formation process \citep[see][for a
review]{Frenk_White:2012}. However, the adequacy of the model remains
controversial in the well observed inner regions of galaxies where the distribution of dark
matter is strongly non-linear \citep[e.g.][]{Walker:2011zu,
  Newman:2013}. Disagreements on these scale are particularly
interesting because they could provide clues to the nature of the dark
matter \citep[e.g.][]{Lovell:2012,Peter:2013,Shao:2013,Zavala:2013}.

A robust prediction of $\Lambda$CDM is that, in the absence of
baryonic effects, the spherically averaged radial density profiles of
dark matter halos of all masses should approximately follow a
universal form, the NFW profile \citep{NFW:1996,NFW:1997}, which
diverges as $r^{-1}$ towards the centre. In galactic halos, baryonic
effects associated with the formation of the galaxy could, in
principle, flatten this central cusp through explosive events produced
by supernovae, as proposed by \cite{Navarro:1996} and seen more
recently in a number of galaxy formation simulations \citep[see][for a
review]{Pontzen:2014}. Energetic arguments suggest that such processes
- if they do indeed occur in nature - should be ineffective in
dwarf galaxies of stellar mass below $10^6$ or $10^7$~M$_\odot$
which would then retain their NFW dark matter cusps
\citep{Penarrubia:2012}. Although difficult to study because of their
intrinsic faintness, the lowest mass dwarf galaxies are thus promising sites for
testing $\Lambda$CDM in the strongly nonlinear regime and learning
about the identity of the dark matter and the effects of baryonic
processes.

\par The most direct way to study the central density structure of a gas-poor 
galaxy is by fitting an equilibrium stellar dynamical model to a large
sample of stars that have high resolution spectroscopy and good
photometry. In recent years, data of the required quality have been
obtained for a number of nearby dwarf spheroidal galaxies (dSphs)
around the Milky Way~\citep{Simon:2007dq,Walker:2008ax} and
M31~\citep{Tollerud:2011mi}. Simple dynamical analyses based on
spherical symmetry and the Jeans equations suffer from degeneracies
which preclude an unambiguous determination of the dark matter
potential ~\citep{Walker:2012td,Strigari:2012gn}. Thus, data for
several dSphs have been shown to be equally consistent with flat central profiles
(cores) ~\citep{Gilmore:2007} or with NFW cusps
~\citep{Strigari:2010un,Jardel:2013oza}. In some cases, one can hope to break degeneracies
by considering higher moments of the line-of-sight
velocity distribution \citep{Richardson:2013lja}.

\par Sculptor, a dSph of stellar mass $\sim 10^7$M$_\odot$ located
$\sim80$~kpc from the Galactic Centre \citep{Lianou_Cole:2013}, is a
particularly interesting case. Modelling using a variety of techniques
but treating the available stellar data as sampled from a single
stellar population and assuming spherical symmetry has shown that the
kinematic data are consistent with an NFW halo potential, but also
allow a core
~\citep{Strigari:2010un,Breddels:2013,Richardson:2013lja}.  These
studies suggest a dark matter halo mass of $\simeq 10^9$~M$_\odot$ for
Sculptor though with rather large uncertainties~\citep{Lokas:2009cp,Strigari:2010un,Breddels:2013}.

\par The data for Sculptor are of sufficient quality that two distinct
stellar populations of differing metallicity can be identified: a
centrally concentrated metal-rich (\MR) population and a more extended
metal-poor (\MP) population~\citep[][B08]{Battaglia:2008jz}. The
presence of two populations makes it possible to carry out more
refined dynamical analyses. Thus, applying the Jeans equations to each
population separately, B08 showed that their data could be fit by a model
in which the orbital distribution of each population is isotropic near
the centre and becomes radially biased in the outer regions.  They
found a best fit for a model potential with a core, but also found the
data to be consistent with an NFW potential. Using Michie-King models
for the stellar distribution function, \cite{Amorisco:2011hb} also
found that while an NFW model provides an acceptable $\chi^2$ fit to
the data, models with a core seem to be preferred. On the other hand,
applying the projected virial theorem,~\cite{Agnello:2012uc} concluded
that it is not possible to fit both the \MR~and the \MP~populations
with a single NFW model.

\par An independent dynamical analysis of Sculptor using a larger
sample of stars was carried out
by~\citet[][WP11]{Walker:2011zu}. Rather than simply separating the
observed stars into two populations according to their estimated
metallicity, they devised a statistical method which fits the full
dataset simultaneously with two constant velocity dispersion,
Plummer-profile populations of differing metallicity, together with a
contaminating Galactic component. They then inserted the half-light
radius and velocity dispersion estimated for each population into the
mass estimator proposed by~\citet{Walker:2009zp}. This allowed them to
infer the mass contained within each half-light radius and thus the
slope of the density profile between the two half-light radii. They
concluded that the slope is flatter than predicted for an NFW profile
at the 99\% c.l.

\par In this study we carry out a new analysis of Sculptor in an
attempt to clarify the conflicting claims in the literature. The
specific statistical question we ask is whether the kinematic and
photometric data for this galaxy exclude potentials of the type
predicted by $\Lambda$CDM.  We re-examine both the B08 and WP11
datasets from a different theoretical perspective and discuss how they
compare. There are both similarities and differences between our
analysis and those that have been undertaken previously.  Like B08 and
\cite{Amorisco:2011hb}, but unlike \cite{Agnello:2012uc} and WP11, we
exploit the full information contained within the line-of-sight
velocity dispersion and photometry profiles. Like
\cite{Amorisco:2011hb}, but unlike B08, we build models based on
distribution functions. Our analysis differs from that of
\cite{Amorisco:2011hb} primarily in that we use a more flexible form
for the distribution function which allows a wider range of energy
distributions and velocity anisotropies for the stars.

We conclude, in agreement with B08 and \cite{Amorisco:2011hb}, that
NFW potentials are {\em not} excluded by the B08 data. 
For our more general models the constraints used by
\cite{Agnello:2012uc} are also no longer sufficient to exclude NFW
potentials. Indeed, the implied peak circular velocity of the Sculptor
dark matter halo and its concentration are consistent with the values
predicted from $\Lambda$CDM simulations.  While an NFW potential gives
an acceptable fit to the Sculptor data, our analysis cannot exclude
potentials with a core; indeed, we find below that a potential of the
form proposed by \citet{Burkert:1995} provides a slightly (though not
significantly) better fit to the B08 data than an NFW profile.
We find similar conclusions when we apply our analysis directly to the WP11 data. 
Thus, our discrepant conclusions reflect differences in analysis methods rather than in observational datasets. 

\par This paper is organized as follows. In Section~\ref{sec:models}
we introduce our model for the stellar distribution function. In
Section~\ref{sec:likelihood} we briefly discuss our methodology for
fitting the theoretical model to the data. In
Section~\ref{sec:results} we present our results and, in
Section~\ref{sec:comparison}, we compare them to previous studies,
highlighting discrepancies where they exist.

\section{Models} 
\label{sec:models}
\par In this section we introduce the dynamical models we use to
interpret the observed stellar populations in Sculptor. We assume each
population to be spherically symmetric and to be in dynamical
equilibrium within a static and spherically symmetric potential
well. These are strong assumptions which should be treated as
approximations. The observed stellar distribution is clearly
non-circular on the sky, and Sculptor orbits within the potential of
the Milky Way, so the effective potential seen by its stars is
time-varying. Some aspects of the effects of flattened potentials on the
dynamical analysis of dSph data are considered by~\citet{Laporte:2014}.

\subsection{Dark matter}
\par For the total mass density profile of the system we adopt two standard models. 
First, an NFW model, 
\be 
\rho(r) = \frac{\rho_s}{x(1+x)^2}, 
\label{eq:NFW}
\ee
with corresponding gravitational potential
\be
\Phi(r) = \Phi_s \left[ 1 - \frac{\ln ( 1 + x)}{x} \right],
\label{eq:NFWpot}
\ee 
where we define $\Phi_s = 4 \pi G \rho_s r_s^2$ and $x = r/r_s$. This
simple model is determined by just two scale parameters, the
characteristic radius, $r_s$, and the characteristic density,
$\rho_s$. Note that we define the potential to be zero at the centre
of the system and to be $\Phi_s$ at infinity. NFW models are often
parametrized in terms of the maximum circular velocity, $V_{max}$, and
the radius, $r_{max}$, at which this is attained. These are related
to $r_s$ and $\Phi_s$ through:

\be
r_{max} = 2.16~r_s; \ \ \ \ \ \ \  V_{max}= 0.465 ~\sqrt{\Phi_s}.
\label{eq:rmaxvmax}
\ee

Our second model is the cored profile proposed by \citet{Burkert:1995}, 
\be 
\rho(r) = \frac{\rho_b}{(1+x_b)(1+x_b^2)}, 
\label{eq:burk}
\ee
where here $x_b = r/r_b$. This profile also has two parameters, the central density $\rho_b$ and the scale radius $r_b$, and we define
$\Phi_b = 4 \pi G \rho_b r_b^2$. The corresponding gravitational potential is 
\begin{multline}
\frac{\Phi(r)}{\Phi_b} = \left(1-\frac{1}{x_b}\right)\frac{\ln (x_b^2
  + 1)}{4} \\ + \left(1 + \frac{1}{x_b}\right) \left(\frac{\tan^{-1}
  (x_b)}{2} - \frac{\ln (x_b + 1)}{2}\right).
\end{multline}
As for the NFW case, we define the potential to be zero at the centre
of the system. With these definitions, as $r \rightarrow \infty$ we
have $\Phi \rightarrow \pi\Phi_b/4$. For the
Burkert model, the maximum of the circular velocity curve and the
radius at which it is attained are related to the other parameters
through \be r_{max} = 3.245~r_b; \ \ \ \ \ \ \ V_{max}=
0.602~\sqrt{\Phi_b}.
\label{eq:rmaxvmax_burk}
\ee

\subsection{Stellar distribution function} 
\par We define the specific energy and specific angular momentum of a
star as $E = v^2/2 + \Phi(r)$ and $J = v r \sin \theta$, respectively,
where $v$ is the modulus of the velocity vector and $\theta$ is the
angle between this vector and the star's position vector relative to
system centre. Given a static and spherically symmetric gravitational
potential well, any positive definite function $f(E,J)$ corresponds to
the phase-space distribution function of some stable, dynamically
mixed and spherically symmetric equilibrium for a stellar
population. In this paper we will consider only models in which the
dependence on $E$ and $J$ is separable, 
\be
f(E,J)=g(J)h(E), 
\label{eq:df}
\ee
with both $g(J)$ and $h(E)$ positive definite and given by simple parametric
forms. It would be easy to build more general, non-separable models as
a superposition of several individually separable components, but we
will not pursue this further here.

\par The stellar density profile and the radial and tangential stellar velocity
dispersion profiles of such models are given by
\bea \rho_\star (r) &=& 2 \pi \int_0^\pi
d \theta \sin \theta \int_0^{v_{esc}} dv v^2 g(J)h(E) \label{eq:rhostar}
\\ \rho_\star \sigma_r^2 (r) &=& 2 \pi \int_0^\pi d \theta \cos^2
\theta \sin \theta \int_0^{v_{esc}} dv v^4 g(J)h(E) \label{eq:sr}
\\ \rho_\star \sigma_t^2 (r) &=& \pi \int_0^\pi d \theta \sin^2 \theta
\sin \theta \int_0^{v_{esc}} dv v^4 g(J)h(E)
                       \label{eq:st}
\eea
where $v_{esc} = \sqrt{ 2 [\Phi_{lim} - \Phi(r) ]}$. Note that with the
definition we are using here, the total velocity dispersion at radius $r$
is
\be
\sigma_{\rm tot}^2(r) = \sigma_r^2(r) + 2\sigma_t^2(r).
\ee
 
\par Eqns~\ref{eq:rhostar},~\ref{eq:sr}, and~\ref{eq:st} can be
combined to give the projected stellar density profile and stellar
line-of-sight velocity dispersion profile at a fixed projected
distance $R$:
\bea 
I_\star(R) &=& 2 \int_0^{\infty}  \rho_\star (r) dz,
\label{eq:IR} \\
I_\star(R) \sigma_{los}^2 (R) &=& 2 \int_0^{\infty}
\rho_\star(r)\frac{ z^2\sigma_r^2 +  R^2\sigma_t^2}{z^2+R^2} dz,
\label{eq:sigmalos} 
\eea
where $r^2 = z^2 + R^2$. 

\par A particularly interesting and simple case occurs when the angular
momentum dependence is taken to be a power law, 
\be 
g(J) = J^b, 
\ee
where $b>-2$ is a constant. For this assumption, the integrals over
$v$ and $\theta$ separate in Eqns~\ref{eq:rhostar},~\ref{eq:sr}
and~\ref{eq:st}, and the ratio of the two velocity dispersions is
independent both of $r$ and of $h(E)$.  The lower limit on $b$ is
required for the $\theta$ integrals to converge for small
$\theta$. For this choice of $g(J)$ the orbital anisotropy of the
stellar population model, usually parametrized as
\be 
\beta(r) = 1 - \sigma_t^2(r)/\sigma_r^2(r), 
\ee 
is independent of radius and depends on $b$ alone, $\beta = - b/2$.
For an isotropic velocity distribution, $\beta=b=0$. For near-radial
orbits $\beta$ is close to unity and $b$ approaches its lower limit of
$-2$, while for near-circular orbits $b$ is very large and positive
while $\beta$ is very large and negative.

\par In this paper we will investigate models where the orbital anisotropy
varies with radius and we therefore need a more general form for $g(J)$.
We consider the function,
\be
g(J) = \left[ \left(\frac{J}{J_\beta}\right)^{\frac{b_0}{\alpha}} + \left(\frac{J}{J_\beta}\right)^{\frac{b_1}{\alpha}} \right]^\alpha,
\label{eq:dfJ}
\ee which interpolates between a power law of index $b_0$ at $J\ll
J_\beta$ and a power law of index $b_1$ at $J\gg J_\beta$. The
parameter $\alpha$ controls the rapidity of the transition between the
two regimes at the characteristic scale, $J_\beta$, which corresponds
to a radius of order $r_\beta=J_\beta/\Phi_\infty^{1/2}$, where
$\Phi_\infty=\Phi_s$ for NFW and $\Phi_\infty = \pi\Phi_b/4$ for the
Burkert case.  In addition, $\alpha$ is required to be positive for
$b_1>b_0$ and to be negative in the opposite case.

For simplicity when comparing with the Sculptor data, we in this paper 
prefer to use a function with fewer free parameters and to assume that
the velocity distribution is isotropic near the centre, as
seems plausible on general theoretical grounds. We therefore set
$|\alpha|=1$ and $b_0=0$, resulting in the simpler expression 
\be
g(J)=
\begin{cases}
    \left[1+(J/J_\beta)^{-b} \right]^{-1}, & \textrm{for } b \le 0 \\
    1 + (J/J_\beta)^b,              & \textrm{for } b > 0.
\end{cases}
\label{eq:dfJs}
\ee The upper and lower cases here correspond to radially and
tangentially biased orbits for large angular momenta,
respectively. Both produce isotropy for small angular momenta and so
also at small radii. This simplified model retains only two
parameters, $J_\beta$ which sets the extent of the inner isotropic
region and $b$ which determines the velocity anisotropy for large
angular momenta.

For the energy distribution, $h(E)$, we have found the following form to be
sufficiently general for our purposes:
\be
h(E) = 
\begin{cases}
N E^a (E^q + E_c^q)^{d/q} (\Phi_{lim} - E)^e & \textrm{for } E < \Phi_{lim}\\
0 & \textrm{for } E\ge \Phi_{lim}, 
\end{cases}
\label{eq:dfE}
\ee
where the restriction $\Phi_{lim}\le \Phi_\infty$ is required because
orbits with $E\ge \Phi_\infty$ are unbound. The normalisation, $N$, in this
expression sets the amplitude of the stellar density profile, while
the exponent $a$ determines the behaviour at small energies, hence as
$r\rightarrow 0$. Comparison with the simple scale-free distribution
functions explored by \cite{White1981} shows that at sufficiently
small radii (where $\Phi \ll \Phi_\infty$, $E\ll E_c$ and $J\ll J_\beta)$)
Eqns~\ref{eq:NFWpot}, ~\ref{eq:dfJ} and~\ref{eq:dfE} imply a
power-law stellar density profile, $\rho_\star\propto r^{-\gamma}$,
where
\be
\gamma = 
\begin{cases}
- a - 3(b_0+1)/2, & \textrm{for } 2a + b_0 < -3 \textrm{ in NFW}\\
- 2(a+b_0) - 3, & \textrm{for } 2a + b_0 < -3 \textrm{ in Burkert}\\
- b_0, & \textrm{for } 2a + b_0 > -3.
\end{cases}
\label{eq:gamma}
\ee
When $2a + b_0 < -3$, the density in the innermost regions
is dominated by stars on orbits which are confined to those regions,
while in the contrary case it is dominated by stars on orbits which extend
well beyond them. Our model for $h(E)$ thus produces a central cusp in
the stellar density profile when Eqn.~\ref{eq:gamma} gives $\gamma>0$.

At somewhat larger energies, $E>E_c$ (hence at radii larger than
$r_c$, where $\Phi(r_c) = E_c$) the density profile steepens to a new
slope, $\gamma^\prime$, which is given by Eqn~\ref{eq:gamma} with $a$
replaced by $a+d$ where we assume $d<0$. The rapidity of the
transition around $r_c$ is controlled by the parameter $q>0$. The
final factor in Eqn~\ref{eq:dfE} allows for truncation of the stellar
density at a radius, $r_{lim}$, defined by $\Phi(r_{lim}) =
\Phi_{lim}$, which is directly analogous to the ``tidal radius'' in
the classic King models for globular clusters \citep{King1966}. The
shape of this cut-off in the profile can be adjusted using the final
parameter, $e$. A special case arises when $\Phi_{lim} = \Phi_\infty$. Then
$r_{lim} \rightarrow \infty$ and the density profile at large radii
becomes a power law of slope $\gamma^{\prime\prime} = e + 3/2 - b_1/2$
(or $\gamma^{\prime\prime} = e + 3/2 - b/2$ for the simpler case of
Eqn~\ref{eq:dfJs}).

As a final remark, we note that when $\Phi_{lim} < \Phi_\infty$, the
constraint $E<\Phi_{lim}$ forces stellar velocities and hence stellar
angular momenta to be small as $r\rightarrow r_{lim}$.  The anisotropy
in this region is thus determined by the form of $g(J)$ for small
rather than large $J$, with the result that $\beta =-b_0/2$ rather
than $-b_1/2$ (i.e. the distribution becomes isotropic again as
$r\rightarrow r_{lim}$ if the simpler parametrisation of
Eqn~\ref{eq:dfJs} is used). This complication does not arise when
$\Phi_{lim} = \Phi_\infty$, in which case $\beta$ is indeed equal to
$-b_1/2$ at large radii ($-b/2$ for the simpler model of
Eqn~\ref{eq:dfJs}).

\section{Data analysis} 
\label{sec:likelihood}
\par In this section we briefly detail the Bayesian analysis methods
that we will apply to the datasets described in the following
sections.

\par For a single stellar population, the model of
Eqns~\ref{eq:df},~\ref{eq:dfJs} and~\ref{eq:dfE} is a function of nine
parameters, $\{N,a,d,q,E_c,\Phi_{lim},e,b,J_\beta\}$. Including the
two parameters that describe the NFW potential, Eqn~\ref{eq:NFWpot}, a
fit to an individual stellar population has 11 free parameters,
whereas a {\em joint} fit to both populations (each of which has
independent distribution function parameters), has a total of 20 free
parameters.

\par We use our theoretical model to fit to the binned velocity
dispersion data. We define the quantities,
\bea 
\chi_{I,p}^2 &=& \sum_{\imath=1}^{n_{p,I}} \frac{ [I_\star(R_\imath) - I_{p}(R_\imath)]^2}{\delta_{p,\imath}^2}, \\
\chi_{\sigma,p}^2 &=& \sum_{\imath=1}^{n_{p,\sigma}} \frac{ [\sigma_{los}(R_\imath) - \sigma_{p}(R_\imath)]^2}{\epsilon_{p,\imath}^2},
\eea 
where the subscript, $p$, denotes a specific
population, either $\MR$ or $\MP$. Additionally
$n_{p,I}$ is the number of data points in the photometric
profile of a population, and $n_{p,\sigma}$ the number of data
points in the velocity dispersion profile of a population,
both of which are measured at projected distance,  $R_\imath$. The
associated measurement uncertainties are $\delta_{p,\imath}$ and
$\epsilon_{p,\imath}$. Using these quantities we define a full
likelihood function of the form,  
\be 
- 2 \ln {\cal L} = \chi_{total}^2 + \textrm{const},
\label{eq:like}
\ee
where 
\be 
\chi_{total}^2 = \chi_{I,\MR}^2 + \chi_{\sigma,\MR}^2 + \chi_{I,\MP}^2
+ \chi_{\sigma,\MP}^2. 
\label{eq:chi2}
\ee
The constant in Eqn~\ref{eq:like} depends on the photometric and
velocity dispersion measurement uncertainties but does not depend on
the distribution function model.  

\par We employ an MCMC algorithm which is an adapted version of the
CosmoMC  code~\citep{Lewis:2002ah}. From this algorithm we can
extract two important quantities: (i) the maximum value of the
likelihood, ${\cal L}_{max}$,  which corresponds to a minimum value of
$\chi_{total}^2$ and to the ``best fit" set of parameters from a given
chain, and (ii)  the posterior probability distribution for each
model parameter. For scans of a large and complex theoretical parameter
space, MCMC algorithms are not necessarily effective at finding the
true value of ${\cal L}_{max}$, so it is important to determine if the
estimated value of ${\cal L}_{max}$ corresponds to a set of
theoretical parameters that provide a statistically good fit to the
data. We thus run several chains from different starting points in the
theoretical parameter space to ensure that the chains find
similar values of ${\cal L}_{max}$ and are thus not burning in at
local maxima in the likelihood.  

\par The fact that we are marginalizing over up to twenty parameters
also means that we must test that the posterior probability
distributions for the model parameters have appropriately
converged. We test for convergence of the posterior probability
distributions in a standard manner by estimating the variance of a
parameter as a weighted sum of the within-chain and between-chain
variance~\citep{Gelman1992}.

\begin{figure*}
\begin{center}
\begin{tabular}{ccc}
{\resizebox{6.0cm}{!}{\includegraphics{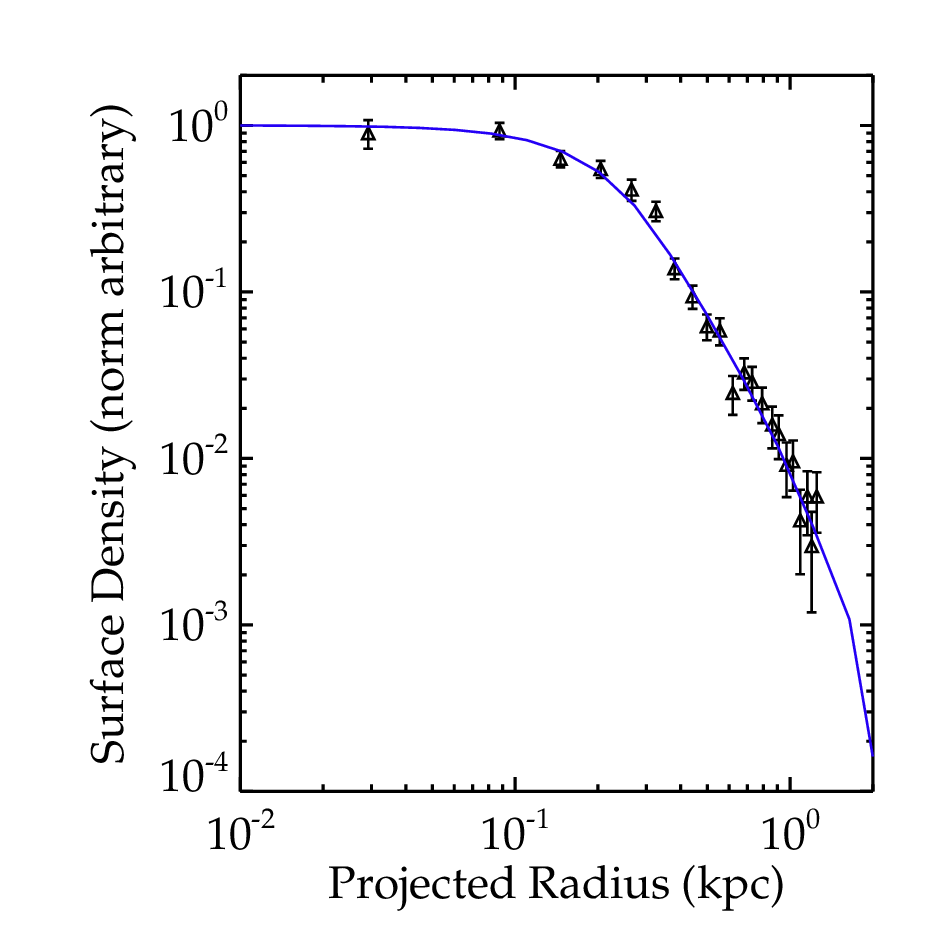}}} &
{\resizebox{6.0cm}{!}{\includegraphics{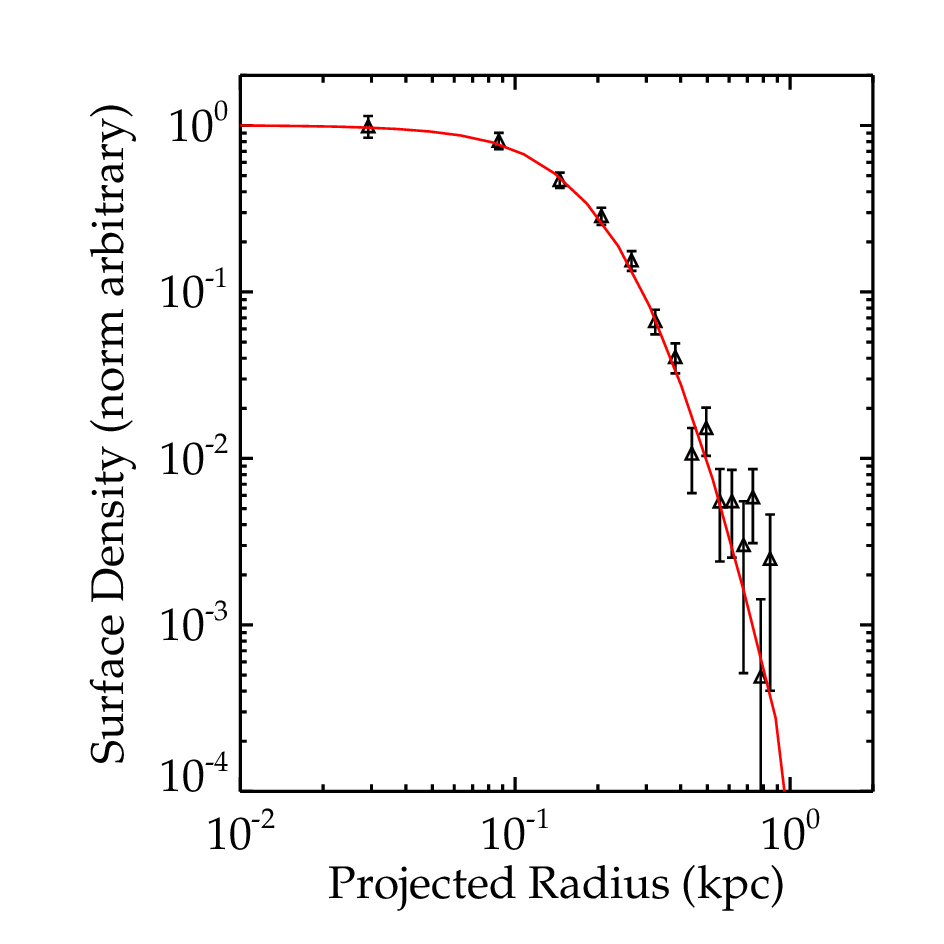}}}& 
{\resizebox{6.0cm}{!}{\includegraphics{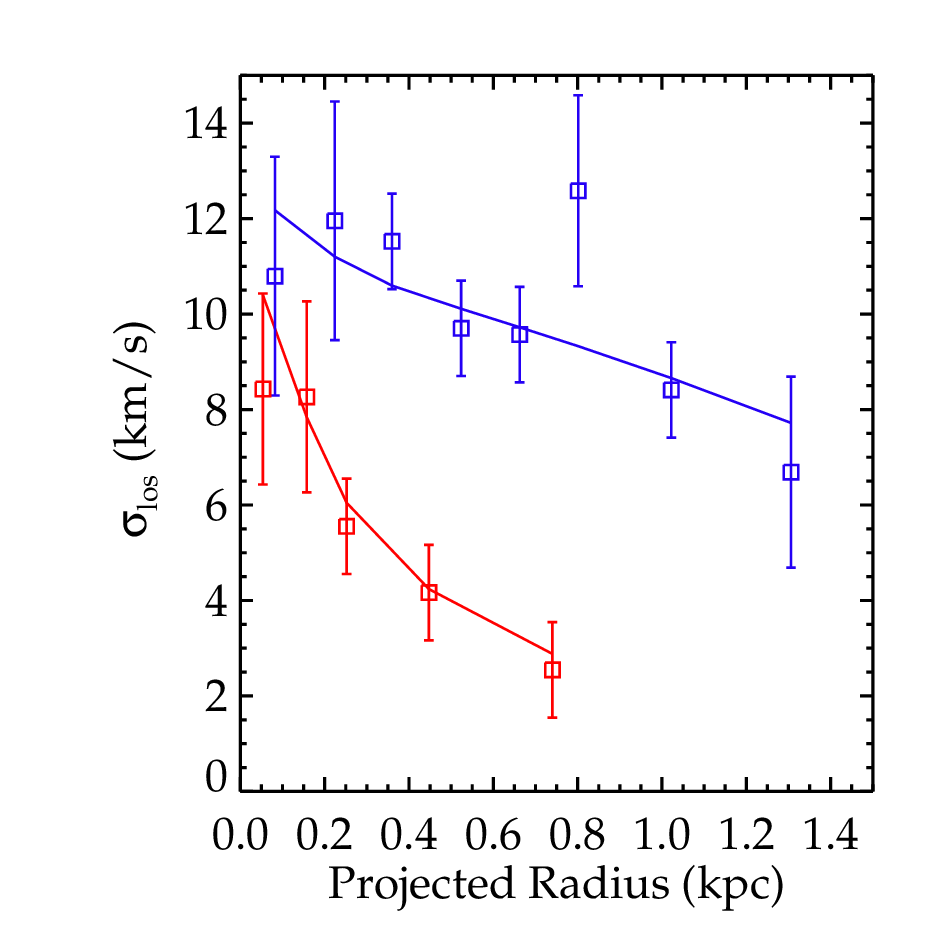}}} \\
\end{tabular}
\end{center}
\caption{Best-fit model of Eqns.~\ref{eq:dfJs} and~\ref{eq:dfE} to
  the~\citet{Battaglia:2008jz} data for the metal-poor and metal-rich
  populations in Sculptor, assuming an NFW potential.  The left and
  middle panels show the surface brightness profiles of the \MP~ and
  \MR~populations respectively; the right panel shows the velocity
  dispersion profile of the \MP~ population (blue line at the top) and
  the \MR~population (red line at the bottom).
  }
\label{fig:profiles}
\end{figure*}

\section{Results}
\label{sec:results}

\par B08 obtained spectroscopy for 470 stars in Sculptor from which
they measured line-of-sight velocities and metallicites derived from
the Calcium triplet lines. They identified two distinct populations
with different metallicity, spatial distribution and kinematics: a
metal rich (\MR) population, defined to have $[Fe/H] >-1.5$, and a
metal poor (\MP) population, defined to have $[Fe/H] < -1.7$. This
clean separation and the large radial coverage of the two populations
make this an attractive sample to analyze using our model distribution
function described in Section~\ref{sec:models}. We fit our
distribution function model to the photometric and velocity dispersion
profiles reported by B08 for each population by performing a
likelihood analysis using the MCMC technique, as described in
Section~\ref{sec:likelihood}. Note that the photometric profiles are obtained 
from a different and much larger population of stars than the velocity dispersion profiles.

\par Since the stellar distribution of Sculptor is elongated on the
sky, B08 give the surface brightness profile of each population as a
function of an ``elliptical radius'' which corresponds to the
projected semi-major axis determined from the
photometry~\citep{Tolstoy2004}. To account for this in the context of
our assumption of spherical symmetry, we take as the radial coordinate
the geometric mean of the major and minor axes, which we expect to
correspond best to the count profile for circular annuli. We also perform the
same scaling on the B08 kinematic data. The
ellipticity of Sculptor is $\epsilon = 0.3$~\citep{Irwin:1995tb}.

\par We fit the full stellar density and velocity dispersion profiles
of the two metallicity populations to the 20-parameter model defined
in Section~\ref{sec:models}, which here assumes an NFW potential, and in
which the velocity anisotropy can vary with radius but is assumed
isotropic at the centre. From the MCMC chains we obtain both the
maximum likelihood value and the posterior probability distribution
for each model parameter. The surface density and velocity dispersion
profiles for a model that has near-maximal likelihood are shown in
Figure~\ref{fig:profiles}.  The count profiles of both stellar
populations exhibit well-defined cores. The \MP~population in this
particular model is isotropic everywhere, while the \MR~population is
isotropic in the centre, has a sharp transition at a scale radius of
$\sim 0.2$~kpc to $\beta \simeq 0.94$ over the range $0.2-1$ kpc, and
then transitions smoothly back to $\beta = 0$ at larger radii in order
to satisfy the boundary conditions at $r_{lim}$. The parameters of
this model are listed in Table~\ref{tab:properties}.

Figure~\ref{fig:profiles} shows that the data for the two metallicty
populations in Sculptor are very well fit by our model, and this
impression is confirmed by the values of $\chi^2$ for the fits: $21.6$
for the \MP~ photometry (left; 23 data points), $8.8$ for the \MR~
photometry (middle; 15 data points), $8.3$ for the \MP~ kinematics
(upper right; 8 data points), and $1.5$ for the MR kinematics (lower
right; 5 data points).  Our analysis therefore demonstrates that the
data are consistent with both populations residing in a single NFW
potential.

\begin{table*}
\begin{center}
  \begin{tabular}{ccccccccccccl}
     Population & $a$ & $d$ & $e$& $E_c$ & $\Phi_{lim}$ & $r_{lim}$ & $b$ & $q$ & $J_\beta$ & $V_{max}$& $r_{max}$ \\ \hline
	MR & 2.0 & -5.3 & 2.5 & 0.16 & 0.45&  1.5 & -9.0 & 6.9 & $8.6 \times 10^{-2}$ & 21 & 1.5 \\  \cline{1-10} 
	MP & 2.4 & -7.9 & 1.1 & 0.17 & 0.60&  3.0 & 0  & 8.2 &--& & \\ \cline{1-10}
  \end{tabular}
  \caption{An example distribution function model that provides a good fit to the Sculptor two-population data. 
  $E_c$ and $\Phi_{lim}$ are in units of $\Phi_s$ and, for the MR population, $J_\beta$ is in units of 
  $r_s \sqrt{\Phi_s}$; $V_{max}$ is in km/s and $r_{max}$ in kpc.}
  \label{tab:properties}
  \end{center}
\end{table*}

\par Consistency with a potential of the NFW form does not, however,
guarantee that the data are consistent with the predictions of the
$\Lambda$CDM model. For this to be the case, the parameters of the NFW
density profile must lie within the theoretically predicted range. The
comparison is most easily carried out in $(V_{max}, r_{max})$ space,
where the maximum circular velocity of the dark matter halo,
$V_{max}$, and the radius, $r_{max}$, at which it is attained are
defined in Eqn.~\ref{eq:rmaxvmax} and are readily measured for
subhalos in high resolution $\Lambda$CDM N-body simulations of
galactic halos.

\par
\begin{figure}
\begin{center}
\begin{tabular}{c}
{\resizebox{7.2cm}{!}{\includegraphics[angle=270]{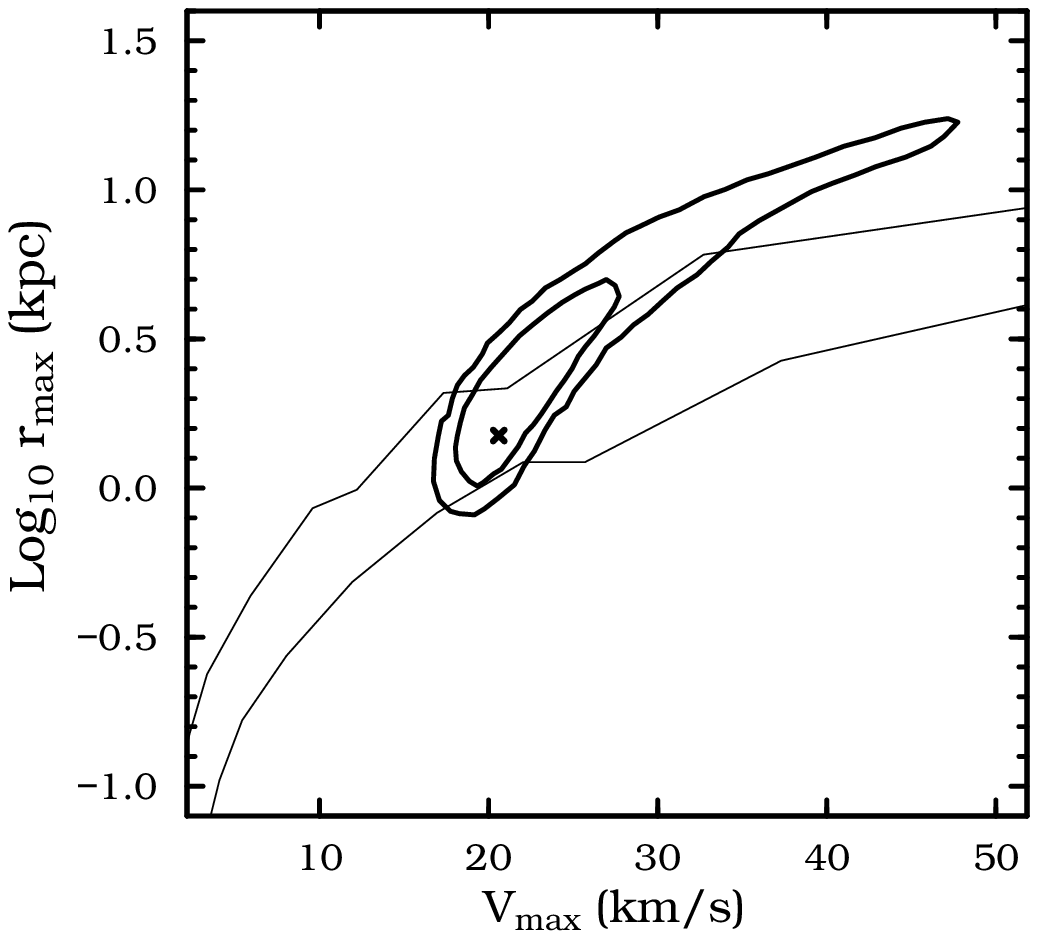}}} \\
\end{tabular}
\end{center}
\caption{The 68\% and 90\% c.l. regions in the $V_{max}-r_{max}$ plane
  for fits of the model of Eqns.~\ref{eq:dfJs} and~\ref{eq:dfE} to
  the two metallicity populations in Sculptor (thick lines). A cross
  indicates the model of Table 1. Thin lines delineate the region which
  contains 90\% of the subhalos from the Aquarius $\Lambda$CDM
  simulations of galactic halos \citep{Springel:2008cc}.  }
\label{fig:vmaxrmax_joint}
\end{figure}

 The region in the $(V_{max}, r_{max})$ plane in which 90\% of the
subhalos in the ``Aquarius'' $\Lambda$CDM simulations of
\cite{Springel:2008cc} lie is shown by the thin lines in
Figure~\ref{fig:vmaxrmax_joint}.  The thick lines show the 68\% and
90\% contours of two-dimensional joint posterior probability
distributions of $V_{max}$ and $r_{max}$ derived from our fits to the
B08 data. These contours overlap well with the theoretically predicted
region, demonstrating that the kinematics of the populations in
Sculptor are fully consistent with expectations in a $\Lambda$CDM
universe. The range of $V_{max}$ allowed by our fits, $\sim
(20-35)$~km/s, is significantly wider than the range estimated in
some previous analyses~\citep{BoylanKolchin:2011de}.

\begin{figure}
\begin{center}
\begin{tabular}{c}
{\resizebox{8.0cm}{!}{\includegraphics[angle=270]{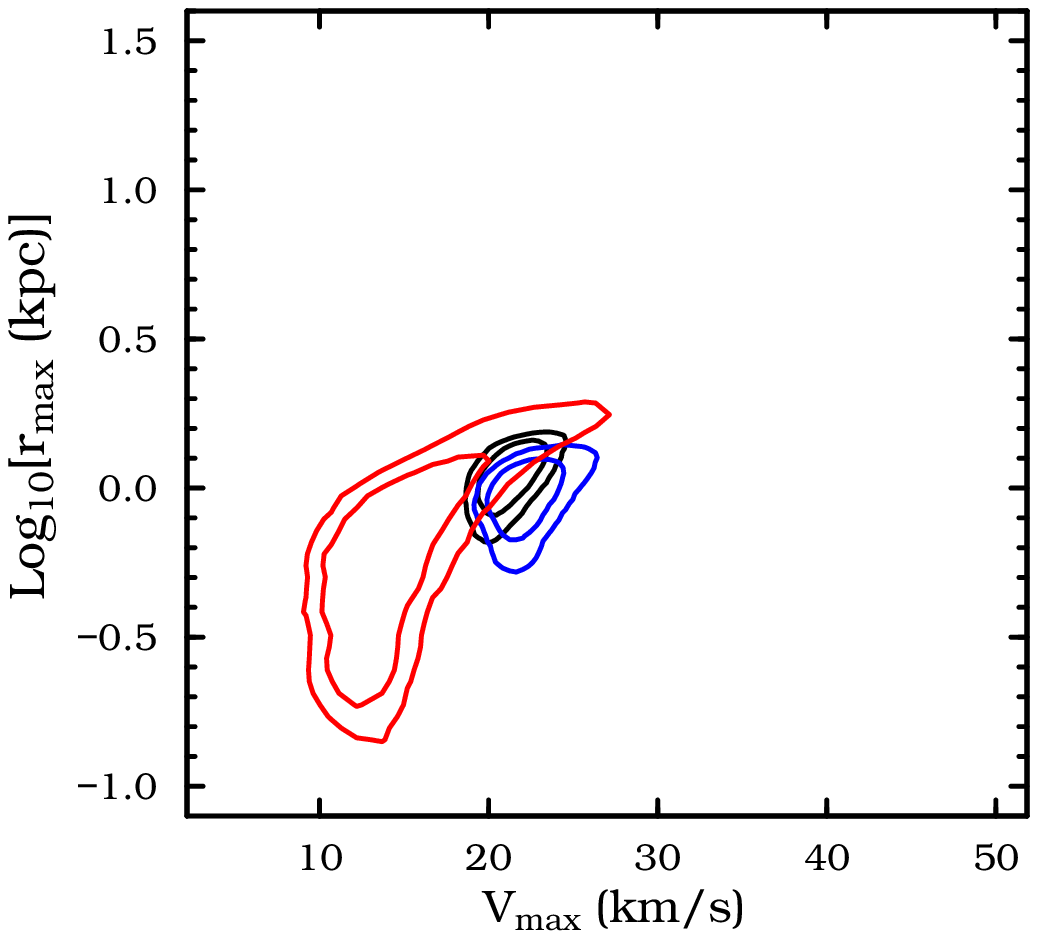}}} \\
\end{tabular}
\end{center}
\caption{68\% and 90\% confidence contours in the $V_{max}-r_{max}$
  plane for independent fits to the two metallicity populations in
  Sculptor assuming a Burkert profile. The red contours are obtained
  by fitting the 11 parameter model of Eqns.~\ref{eq:dfJs}
  and~\ref{eq:dfE} to the~B08~\MR~data, while the blue contours are
  obtained from fitting the same 11 parameter model to
  the~B08~\MP~data. These contour sets can be compared with
  Figure~\ref{fig:vmaxrmax_individual} for the NFW profile. The black
  contours are the results of a joint fit to the~\MR~and~\MP~data as
  shown for the NFW profile in Figure~\ref{fig:vmaxrmax_joint}.}
\label{fig:burk_all}
\end{figure}

To compare the quality of our NFW fits to that obtained
  for a cored potential, Figure~\ref{fig:burk_all} presents results
  from a joint analysis of the two populations using a Burkert
  profile. In this case, we also have 20 parameters, but we replace
  $(\rho_s, r_s)$ by $(\rho_b, r_b)$. The best-fit $V_{max}-r_{max}$
  values in this figure are almost identical to those found in the NFW
  case, but the constraints are considerably tighter, reflecting the
  much more sharply defined characteristic scale of the cored
  potential.  The total $\chi^2$ for the best fit is $39.3$ in the
  Burkert case which is slightly but not significantly smaller than
  the value of 40.2 which we found for the best-fitting NFW potential.

  Given the similar quality of the fits for the two profiles, we
  consider the more general question of whether we can expect to
  distinguish them with kinematic and photometric datasets similar to
  those of B08. We start from distribution function parameters
  resembling those of our best Burkert fit to B08, and generate mock
  photometric and kinematic datasets with similar size and
  uncertainties to B08. We consider several values for the Burkert
  scale radius and scale density, chosen to give dispersion and count
  profiles similar to those observed. The largest value for the scale
  radius we consider is $r_b = 1$ kpc.  We fit these mock data to
  Burkert and NFW models, comparing the best fit $\chi^2$ values in
  the two cases. Even for $r_b = 1$ kpc, we find NFW models with very
  similar $\chi^2$ to the best fitting Burkert model. The latter
  always has parameters very close to the input values, showing our
  procedure to be approximately unbiased. This test implies that the
  B08 data samples are not large enough to be able to distinguish
  between Burkert and NFW potentials on the basis of count and
  velocity dispersion profiles. We will return to the issue of
  distinguishing between these two models below when discussing the
  WP11 data.  

\section{Comparison to previous results} 
\label{sec:comparison}

\par Several previous studies have constrained the potential in
Sculptor by splitting its stars into high and low metallicity
populations and requiring each separately to be in equilibrium. In
this section we compare our results with these earlier analyses.

\subsection{Battaglia et al. (2008)}

\par In the first dynamical analysis to separate the two populations
in Sculptor, B08 applied the Jeans equation to each population
individually, fitting the star count profiles to standard forms (Plummer
for the metal-rich, Sersic for the metal-poor) and then predicting the
velocity dispersion profiles for various assumed potentials and
anisotropy profiles. They found that their data were consistent with
both NFW and core potentials, but required radially biased orbits for
both populations.

Our results largely corroborate the conclusions reached by B08.  Our
use of a flexible stellar distribution function removes the need to
assume standard forms for the count and anisotropy profiles and
ensures that the resulting model is physically realisable.  As a
result, our best-fit model has a smaller $\chi^2$ than the model of
B08. Like B08, we find that radially-biased orbits are required at
large radius for the metal-rich (although not for the metal-poor)
population. 

\subsection{Amorisco \& Evans (2012)}

Of the previous studies, that of ~\cite{Amorisco:2011hb} is most
similar to our own. They also assumed separable distribution functions
for the two populations and fit predicted counts and velocity
dispersion profiles to the data of B08. They considered both NFW
potentials and pseudo-isothermal potentials with a core.  Although the
form they assumed for their distribution functions is considerably
less flexible than our own, the resulting best fit for an NFW
potential is similar to ours, shown in Fig.~\ref{fig:profiles}, and
has a $\chi^2$ value which is clearly insufficient to exclude the
model. The best fit for the core case also looks similar (compare
their figures 9 and 10). Nevertheless, its $\chi^2$ value is
sufficiently smaller that a likelihood ratio test clearly prefers it
over an NFW potential.

As \cite{Amorisco:2011hb} note, the preference for a core potential
over a cuspy one is driven in their analysis by its lower prediction
for the innermost points of the count profiles and, to a lesser
extent, by a somewhat larger predicted difference in velocity
dispersion between the two populations. With our more flexible
distribution function model, the count discrepancy at small radius
disappears for the NFW potential and the difference in velocity
dispersions between the \MP~and~\MR~populations is slightly enhanced
(see Fig.~\ref{fig:profiles}), leading to a fit of very similar
quality to that found by ~\cite{Amorisco:2011hb} for their core
potential. A final difference with \cite{Amorisco:2011hb} is that
their NFW fit required a halo concentration which is lower than
expected in $\Lambda$CDM. With our distribution function model, this
problem has disappeared.

\subsection{Agnello \& Evans (2012)}

\par~\cite{Agnello:2012uc} applied the projected virial theorem
separately to the two populations identified by B08 assuming that they
reside in an NFW potential and have Plummer-law surface brightness
profiles. With these assumptions, the observational data for each
population imply a relation between $V_{max}$ and $r_{max}$ for its
halo. They then show that the regions of the ($V_{max}$, $r_{max}$)
plane allowed at 2$\sigma$ by the \MR~and~\MP~data do not
overlap. They therefore conclude that no single NFW potential can
accommodate both populations.

\begin{figure}
\begin{center}
\begin{tabular}{c}
{\resizebox{7.7cm}{!}{\includegraphics[angle=270]{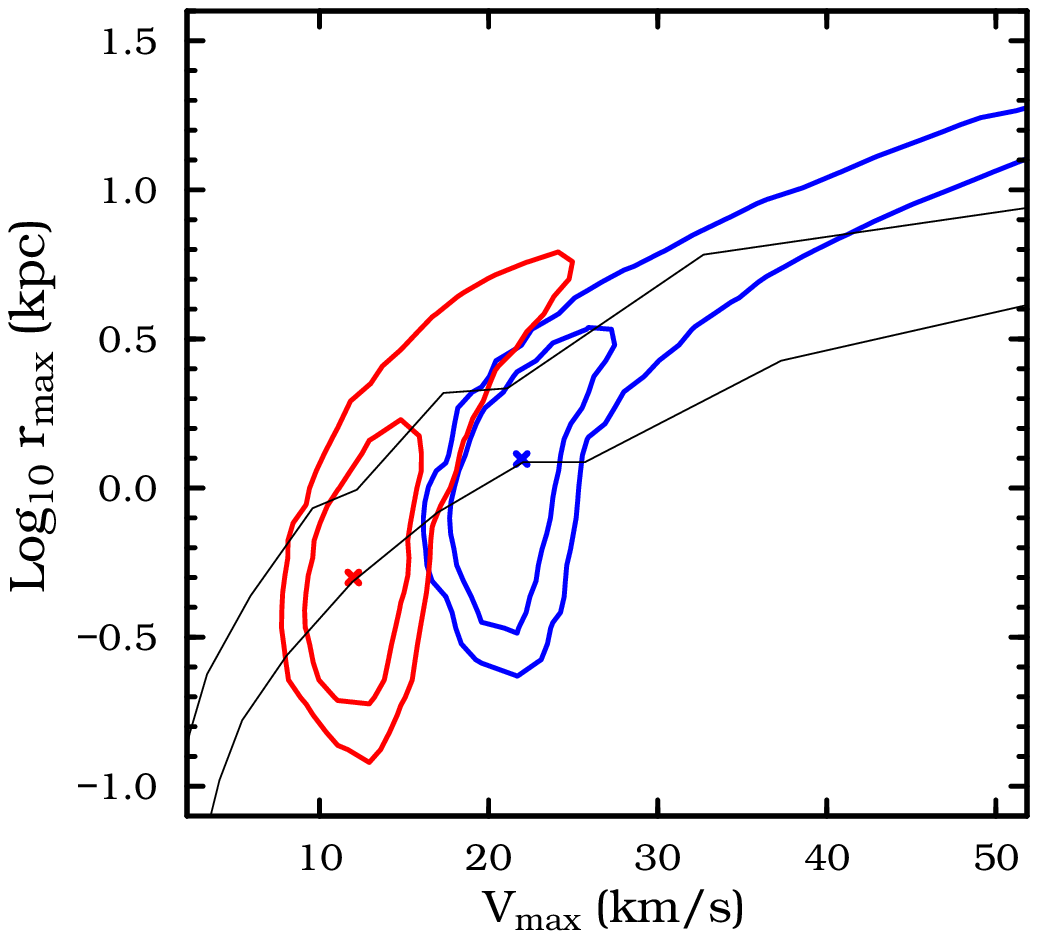}}} \\
\end{tabular}
\end{center}
\caption{The 68\% and 90\% c.l. contours in the $V_{max}-r_{max}$
  plane for {\em independent} fits to the two metallicity populations
  in Sculptor. The red contours are obtained from fitting the 11
  parameter model of Eqns.~\ref{eq:dfJs} and~\ref{eq:dfE} to
  the~B08~\MR~data, while the blue contours are obtained from fitting
  the 11 parameter model to the~B08~\MP~data. Crosses in each case
  indicate the best-fit model. }
\label{fig:vmaxrmax_individual}
\end{figure}

B08 noted that the metal-poor population in Sculptor is poorly fit by
a Plummer model. By applying our procedures to the \MR~and~\MP~data
separately, we can perform an analysis analogous to that of
\cite{Agnello:2012uc}. In Figure~\ref{fig:vmaxrmax_individual} we show
the two-dimensional joint posterior probability distributions of
$(V_{max}$, $r_{max})$. We indeed reproduce an offset similar to that
seen by \cite{Agnello:2012uc}. However, the greater freedom afforded by
our relaxation of the Plummer-law assumption, allowing instead any
profile consistent with the observed counts, widens the confidence
regions so that they are no longer exclusive.  The two distributions
are marginally consistent with each other for $V_{max} \sim
(15-25)$~km/s and $r_{max} \sim 1$~kpc which, not suprisingly, is the
region also picked out by our single potential model. As before, these
parameters are consistent with the $\Lambda$CDM predictions of
\citet{Springel:2008cc}.

\subsection{Walker \& Penarrubia (2011)} 

\par WP11 analyzed a sample of 1497 stars with spectroscopy from which
they derive line-of-sight velocities and an Mg index which they take
as a proxy for metallicity. In practice, we are able to work with the subsample of
1307 stars for which the public online dataset lists a membership
probability value. WP11 use MCMC techniques to map the
parameter distributions for a 3-component model of these data. The
metal-rich and metal-poor populations are each represented by a
circularly symmetric Plummer profile, with Gaussian velocity and
metallicity distributions independent of radius
(note that these assumptions are not consistent with the B08 data), while 
the contaminating Galactic foreground is taken to be spatially uniform
with broader distributions of velocity and metallicity. They 
insert the half-light radius and velocity dispersion estimated by this analysis for each
population into the mass estimator proposed by~\citet{Walker:2009zp}:
\begin{equation}
M_{h}=M(R_h) = 2.5 \langle \sigma_{los}^2 \rangle R_h/G, 
\label{eq:wolf}
\end{equation}
which gives the mass, $M_{h}$, inside a sphere with radius equal to
the projected half-light radius, $R_h$, in terms of the measured
velocity dispersion, $\sigma_{los}^2$, and $R_h$\footnote{This
  estimator is constructed to be only weakly sensitive to the details
  of the density and velocity anisotropy profiles \citep[see
  also][]{Wolf:2009tu}}. The derived increase in estimated mass
between the two values of $R_h$ appears too large to be consistent
with an NFW profile and is close to that expected for a constant density
core. WP11 conclude that NFW is excluded at the 99\% c.l.

\begin{figure}
\begin{center}
\begin{tabular}{c}
{\resizebox{7.0cm}{!}{\includegraphics[angle=270]{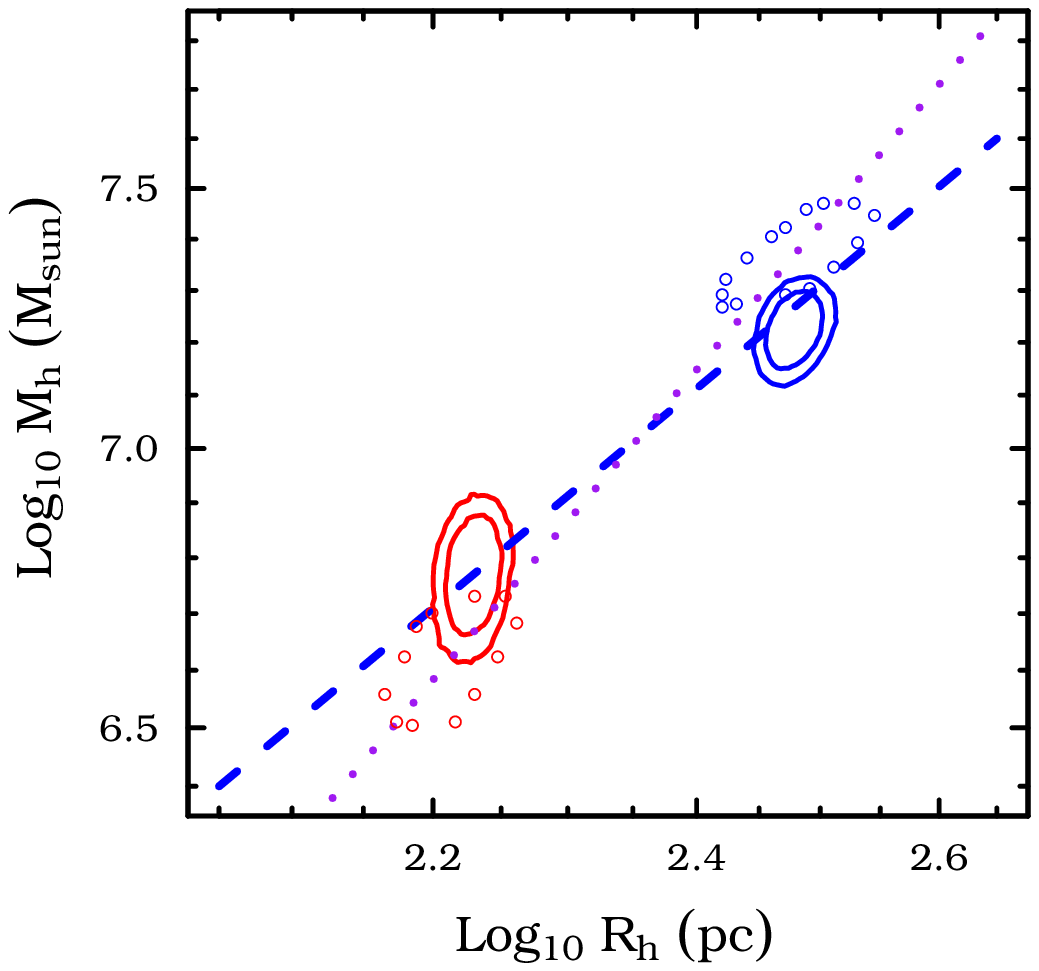}}} \\
\end{tabular}
\end{center}
\caption{Constraints from  WP11 and from the models of this paper in
  the ($R_h, M_h$) plane. The two straight lines and the contours traced by open circles are
taken directly from Figure 10 of WP11  and indicate $M\propto r^\alpha$ with
$\alpha=2, 3$ and the 50\% c.l. regions given by their MCMC analysis 
for the parameters of the two underlying populations.  
For comparison, the solid contours show 68\% and 90\% c.l. regions 
from our own MCMC chains constrained by the B08 data and assuming both
populations to be in equilibrium within a single NFW potential.
}
\label{fig:wp}
\end{figure}

This conclusion is incompatible with our conclusion derived above, 
based on the B08 data. In Fig.~\ref{fig:wp} we show the results of WP11 in the ($R_h$,
$M_h$) plane, together with lines corresponding to $M\propto
r^\gamma$, with $\gamma=2$ and $\gamma=3$. Clearly, these results
agree much better with the dotted line representing a core than with
the dashed line representing an NFW cusp. Our distribution function
based MCMC analysis allows us to reconstruct $R_h$ and $M_h$ for all
models consistent with the B08 data and residing in an NFW
potential. Solid red and blue contours in Fig.~\ref{fig:wp} give the
68\% and 90\% confidence regions for the metal-rich and metal-poor
populations respectively.  As expected, the centre points of these
contours define a slightly shallower slope than the dashed line since
$\gamma=2$ only in the innermost regions of an NFW profile. The
half-light radii found for the \MR~and \MP~ populations in the two
analyses agree well but there is an offset in the preferred $M_h$
values, although the contours do overlap. Hence, fitting our models to
the B08 data has resulted in lower velocity dispersions for the
\MP~and higher velocity dispersions for the \MR~ population than estimated by WP11
from their own data.

\begin{figure*}
\begin{tabular}{cc}
{\resizebox{9.3cm}{!}{\includegraphics[angle=0]{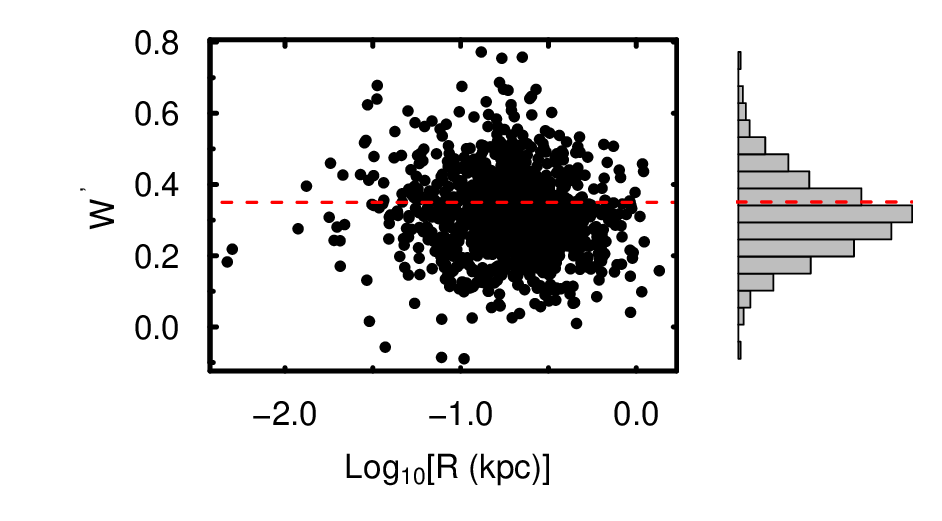}}} & 
{\resizebox{6.5cm}{!}{\includegraphics[angle=0]{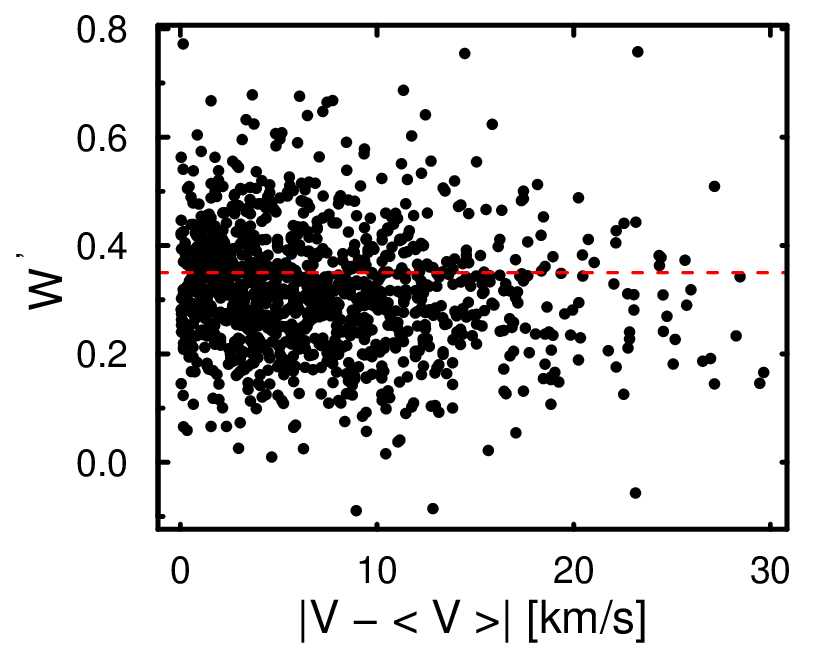}}} \\
\end{tabular}
\caption{Reduced Mg index, $W^\prime$, versus radius (left) and
  velocity (right) for stars considered to be probable Sculptor
  members by WP11.  
  The histogram plotted vertically (center) 
  shows the distribution of $W^\prime$ for the sample as a whole.
  The red dashed line defines the cut which maximally separates the radial distributions of metal-rich and metal-poor 
  populations. 
\label{fig:MgvsR}}
\end{figure*}

To determine whether the WP11 data favor a cored or cusped 
halo when analyzed using our procedures, we construct binned photometric 
and velocity dispersion profiles directly from the WP11 data. Figure~\ref{fig:MgvsR}
shows the metallicity as a function of both radius and velocity for 
the 1160 stars in the tables published by WP11 that have high quality data and are assigned 
a membership probability $> 99$\%.  The W$^\prime$ 
distribution of the member stars is unimodel, with no obvious indication of two distinct 
populations. To separate the stars by metallicity into two populations with maximally 
distinct spatial distributions, we split them at a given value of $W^\prime$ in the left panel of 
Fig.~\ref{fig:MgvsR}, and then perform a KS test to determine 
the significance of the difference in radial distribution between the two populations. The
value of $W^\prime$ which minimizes 
the KS p-value then defines the~\MR~and~\MP~populations with the most 
significantly distinct spatial distributions. 

Following this procedure, we find a minimum p-value of $1.4 \times 10^{-7}$ at
$W^\prime = 0.35$. This cut gives $763$~\MP~members with $W^\prime \le 0.35$, and $397$~\MR~ 
members with $W^\prime > 0.35$. The left panel of Fig.~\ref{fig:MgvsR} shows that, with this cut, the fraction of MR stars at large
radius is noticeably smaller than for the MP stars. Including all 1307 WP11 stars with a membership probability,
regardless of its value, increases the p-value by nearly two orders of magnitude, but still gives an optimal separation at about the same W' value,
and with a similar ratio~\MR~to~\MP~stars.  
For comparison, in their statistical separation, WP11 found that $53$\% of the Sculptor member stars belong to 
the underlying "true"~\MR~population, and the remainder to the~\MP~population. 

The velocity dispersion and the photometry profiles of the two populations are shown in 
Figure~\ref{fig:profiles_wp_mrmp}. 
In comparison to the B08 data in Figure~\ref{fig:profiles}, the WP11 data cover a narrower range of radii
but have higher signal to noise for the velocity dispersion measurements. 
For all four profiles, we have corrected the raw numbers using the radial selection 
function as suggested in WP11. 
We find the half-light radii for the~\MR~and~\MP~populations to be 0.18 and 0.22 kpc, respectively.
The~\MR~half-light radius is in good agreement with that derived by 
WP11, while the~\MP~half-light radius is significantly smaller than the value of 0.30 kpc derived by WP11. 
The velocity dispersion profile of the~\MR~component derived from the WP11 data does not show 
a steep decline at large radii of the kind seen in the B08 data. 

\begin{figure}
\begin{tabular}{c}
{\resizebox{7.7cm}{!}{\includegraphics[angle=270]{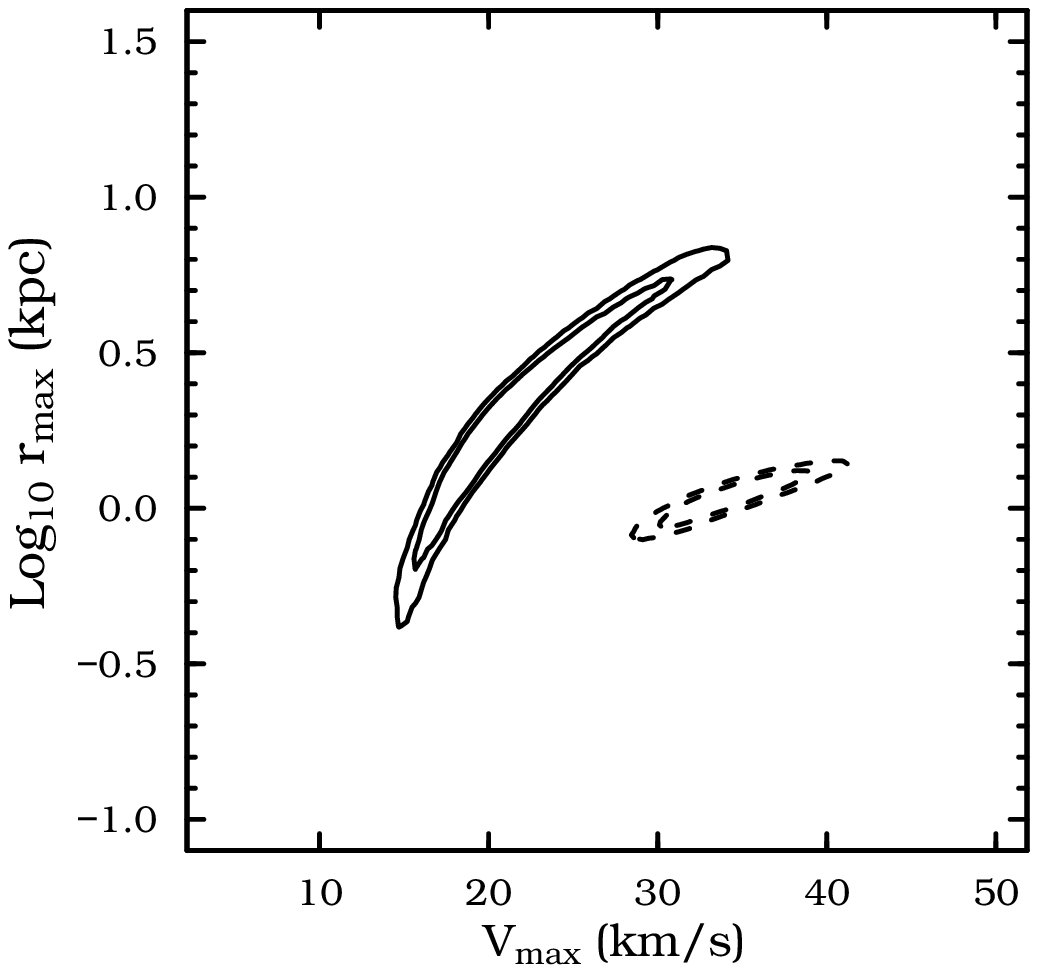}}} \\
\end{tabular}
\caption{The 68\% and 90\% c.l. contours in the $V_{max}-r_{max}$
  plane for {\em joint} fits to the two metallicity populations
  in Sculptor using the WP11 data split according to the cut in Fig.6. Solid contours are for an NFW and dashed are for a Burkert potential.}
\label{fig:vmaxrmax_individual_wp}
\end{figure}  
 
An MCMC analysis of the WP11 results in allowed regions in the $V_{max}-r_{max}$ parameter space
shown in Figure~\ref{fig:vmaxrmax_individual_wp} for both Burkert and NFW profiles. For the NFW profile, 
the derived values of $V_{max}$ are in good agreement with the values derived from the B08 data, 
while for the Burkert profile the central $V_{max}$ is larger, peaking at $\sim 30$ km/s. The corresponding 
best fitting photometry and velocity dispersion profiles are shown in 
Figure~\ref{fig:profiles_wp_mrmp}. The $\chi^2$ 
for the best fitting profiles for both joint and individual fits are given in Table~\ref{tab:WPbestfits}. 
Note that in both cases we are fitting 34 data points to a 20 parameter model. so the $\chi^2$ values we find 
indicate fully acceptable fits and show a slight preference for NFW over Burkert.

\begin{figure*}
\begin{tabular}{cccc}
{\resizebox{6.0cm}{!}{\includegraphics[angle=0]{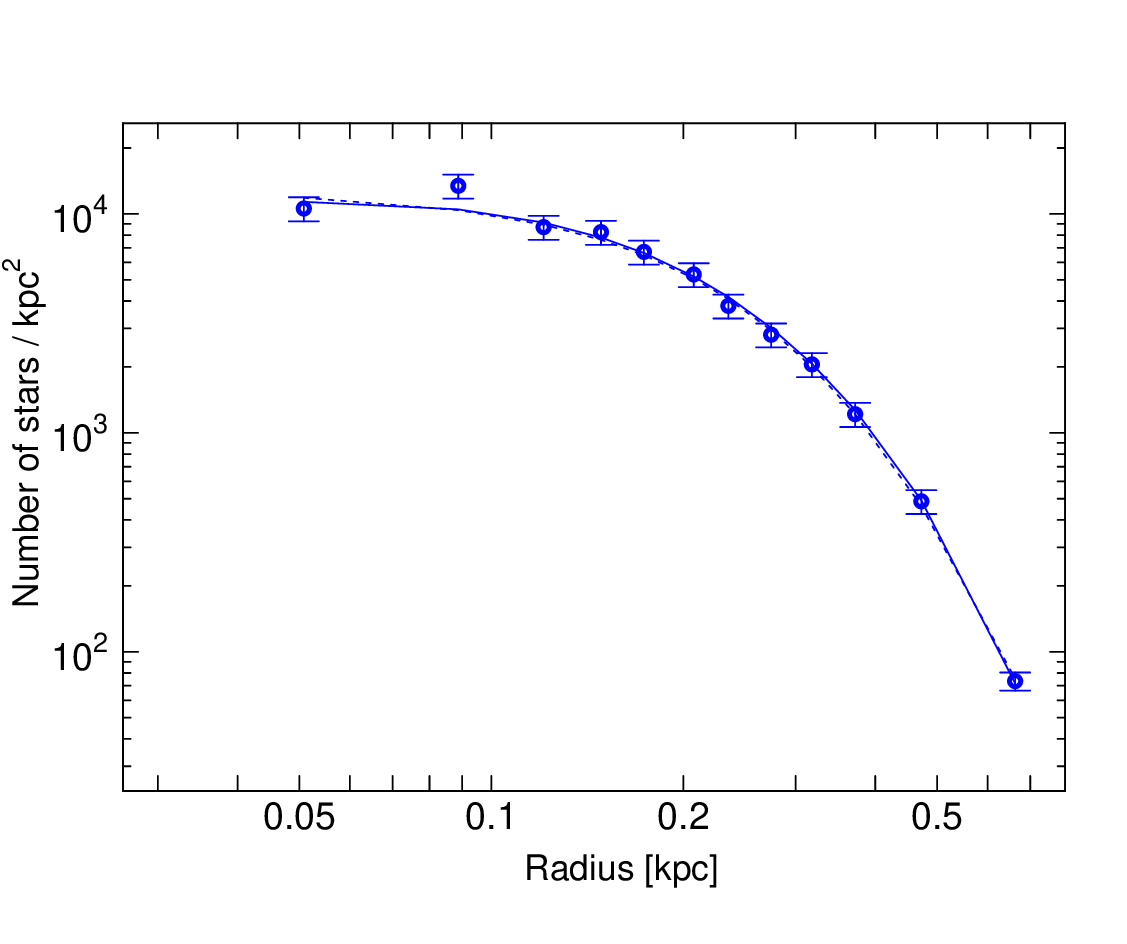}}}& 
{\resizebox{6.0cm}{!}{\includegraphics{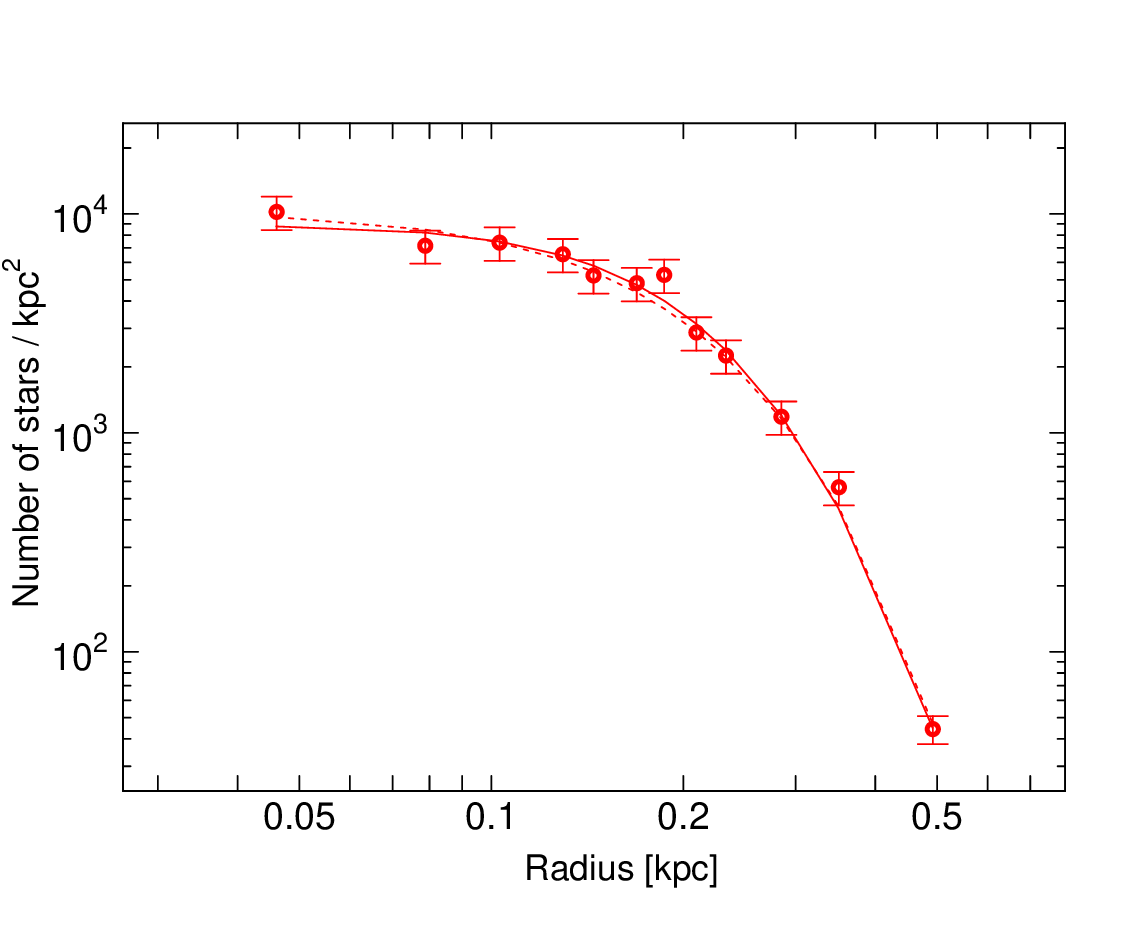}}}& 
{\resizebox{6.0cm}{!}{\includegraphics{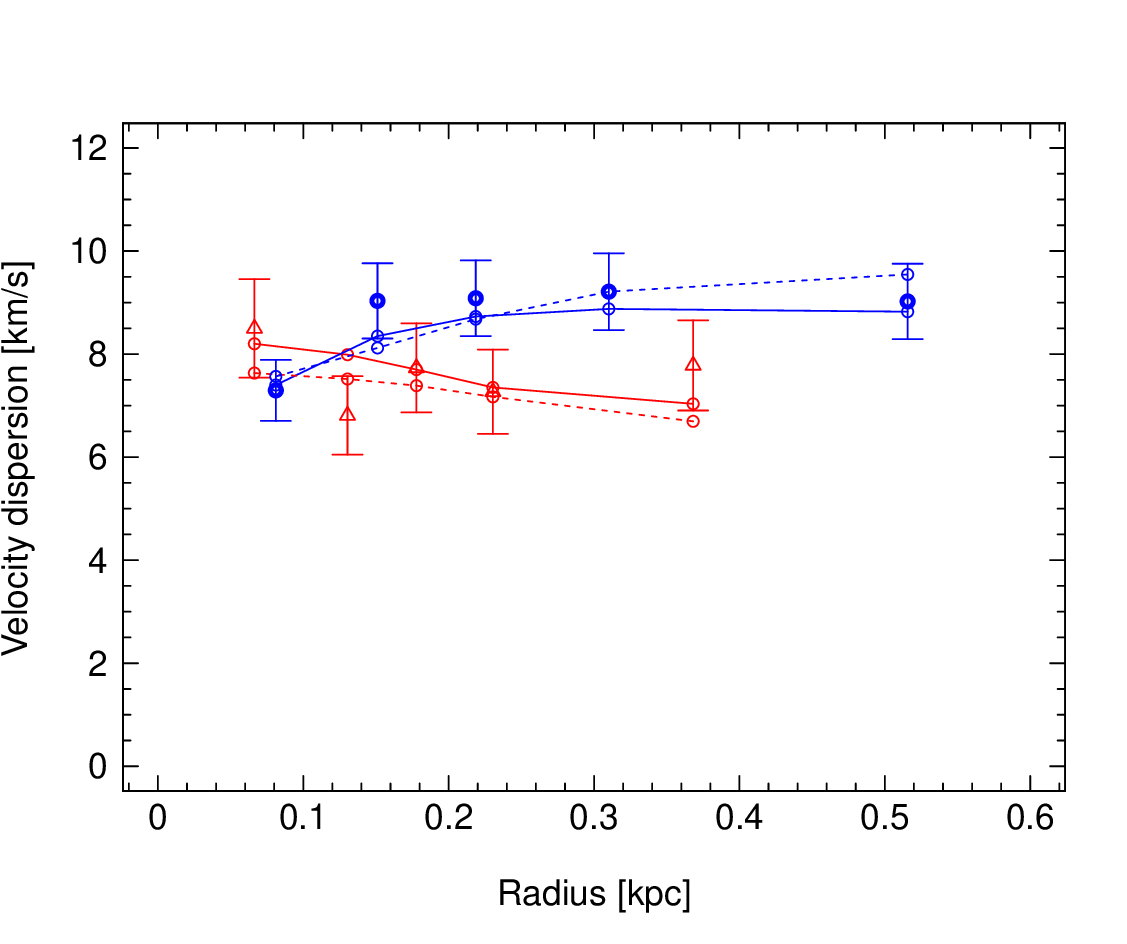}}} \\
\end{tabular}
\caption{Results of {\it joint} fits to the ~\MP~and~\MR~ photometry and kinematics, using the WP binned data. In each panel, 
the solid line assumes an NFW, and the dashed line a Burkert potential. 
 }
\label{fig:profiles_wp_mrmp}
\end{figure*}

\begin{table}
\caption{Total $\chi^2$ values for the best-fit to the photometry and kinematic profiles using the WP11 data.}
\begin{tabular}{lccccc}
\hline 
    &\multicolumn{2}{c}{NFW}&& \multicolumn{2}{c}{Burkert}\\
     \cline{2-3}
    \cline{5-6} 
    &photometry&dispersion&&photometry&dispersion\\
    \hline
        	MR&5.5&3.9&&6.0&2.8\\
    \hline
        MP&5.0&1.2&&5.5&3.2\\
    \hline
\end{tabular}
\label{tab:WPbestfits}
\end{table}

\par We thus conclude that when analyzed with the methods of this paper the WP11 data
do not favour a core over a cusp. This result is in disagreement with that
of WP11, who find NFW profiles to be ruled out with $> 99$\% confidence. At this point we are unable to 
determine why our conclusions differ from those of WP11. 
Our dynamical analysis is based on
self-consistent distribution function models with considerable
flexibility which can be exploited to obtain good simultaneous fits to
the photometric and kinematic data. In contrast, WP11 use simple,
constant velocity dispersion Plummer-law models which are inconsistent
in detail with the Sculptor data. However, since these are used only
to estimate the half-light radii and total velocity dispersions of the
two populations, it is unclear whether this inconsistency can
significantly bias their results. A second difference is that our
likelihood analysis explicitly separates Sculptor stars into two
populations based on the directly measured quantity W' and rejects
{\it ab initio} the small number of stars which are not high
probability members, whereas WP11 treat W' as an indicator
of the relative probability of belonging to
one or the other of two assumed underlying
populations, and also treat background rejection
probabilistically. However, at the present time it is unclear whether
these differences can account for the difference in our conclusions.
Fig.~\ref{fig:profiles_wp_mrmp} does appear to demonstrate that our model can reproduce the
WP11 data very well in an NFW potential.  

\section{Conclusions} 

\par In this paper we have presented a new distribution function
based framework to study stellar populations in equilibrium within a
spherical dark matter potential well. We use it to study the two 
metallicity populations in the Sculptor dwarf spheroidal galaxy, in
particular, to explore the controversial question of whether their
properties exclude a cuspy profile of the kind expected in the
$\Lambda$CDM cosmology. 

The family of distribution functions we consider gives substantially
more freedom than the models assumed in previous studies and, as a
result, leads to a weakening of the constraints implied by the
observations.  Although in the absence of any prior on the shape of
the inner potential, we concur with previous studies that the Sculptor
data prefer a shallower profile than NFW, we find this preference to
be far too weak to exclude the cosmological prediction. Indeed, in a
$\chi^2$ sense, we are able to find equilibrium models which are 
a good fit to the datasets of both B08 and WP11 within an NFW
potential with parameters that are fully consistent with $\Lambda$CDM.

Since the inner structure of dwarf galaxies appears at present as one
of the few significant challenges to the standard cosmological
paradigm it is unsurprising that considerable attention has been focused on
measuring this structure precisely. Unfortunately, the problem is
underconstrained by currently available data, given the considerable
freedom inherent in the equations of stellar dynamics. The analysis in
this paper, while comparatively general, still makes at least two
major assumptions which are known to be incorrect: dwarf spheroidal
galaxies are clearly not spherically symmetric and their orbits within
the Milky Way's potential ensure that most cannot be static systems in
equilibrium. Further theoretical progress will require these
shortcomings to be addressed (See  Zhu et al (2016) for a recent study of 
Sculptor which relaxes the assumption that the stelllar distribution is spherical, 
while continuing to assume a spherical potential.)
Further observational progress may be
achieved by reducing the statistical and measurement uncertainties
and, in the more distant future, by increasing the phase-space
coverage through measurement of internal proper motions.

\section*{Acknowledgements}

We would like to thank Chervin Laporte for his critical comments on an
earlier version of this work which prompted us to develop the
distribution function methods used in this paper. We also thank Matt Walker
for critical and constructive comments on the paper. Much of the theoretical modelling 
framework for this paper was developed at the Aspen Center for Physics, which is supported by
National Science Foundation grant PHY-1066293. LES acknowledges support 
from National Science Foundation grant PHY-1522717. This material is
based upon work supported by the National Science Foundation under
Grant No. CNS-0723054.  CSF acknowledges ERC Advanced Investigator
Grant COSMIWAY and SDMW ERC Advanced Investigator Grant GALFORMOD.
This work was supported in part by an STFC rolling grant to the ICC.
This work used the DiRAC Data Centric system at Durham University, operated by the Institute for 
Computational Cosmology on behalf of the STFC DiRAC HPC Facility (www.dirac.ac.uk). This equipment 
was funded by BIS National E-infrastructure capital grant ST/K00042X/1, STFC capital grant ST/H008519/1, 
and STFC DiRAC Operations grant ST/K003267/1 and Durham University. DiRAC is part of the National E-infrastructure. 

\bibstyle{mn2e} 

\newpage 
\newpage

\end{document}